\documentclass[aps,preprint,superscriptaddress]{revtex4}

\usepackage{epsfig}
\usepackage{amsmath}

\begin{document}
\bibliographystyle{apsrev}


\title{Phase diffusion as a model for coherent suppression of tunneling
in the presence of noise}



\author{J.~Grondalski}
\email[]{jcat@unm.edu}   
\homepage[]{http://panda30.phys.unm.edu/Deutsch/Homepage.html}

\affiliation{Albuquerque High Performance Computing Center,\\
University of New Mexico, Albuquerque, NM}

\affiliation{Department of Physics and Astronomy,\\
University of New Mexico, Albuquerque, NM}

\author{P.~M.~Alsing}

\affiliation{Albuquerque High Performance Computing Center,\\
University of New Mexico, Albuquerque, NM}

\author{I.~H.~Deutsch}

\affiliation{Department of Physics and Astronomy,\\
University of New Mexico, Albuquerque, NM}


\date{\today}

\begin{abstract}

We study the stabilization of coherent suppression of tunneling in a
driven double-well system subject to random periodic $\delta-$function
``kicks''.  We model dissipation due to this stochastic process as a phase
diffusion process for an effective two-level system and derive a
corresponding set of Bloch equations with phase damping terms that agree
with the periodically kicked system at discrete times.  We demonstrate
that the ability of noise to localize the system on either side of the
double-well potenital arises from overdamping of the phase of oscillation
and not from any cooperative effect between the noise and the driving
field.  The model is investigated with a square wave drive, which has
qualitatively similar features to the widely studied cosinusoidal drive,
but has the additional advantage of allowing one to derive exact analytic
expressions.

\end{abstract}



\maketitle


\section{Introduction}

Coherent suppression of tunneling is a localization effect that occurs
when a potential with
multiple minima is exposed to a periodic drive in a specific parameter
regime \cite{Dun86,Gro91,Hol92}.  In their original paper, Grossman et.
al. \cite{Gro91} examined
this phenomena in the context of a quartic double-well potential driven by
a strong cosinusoidal force.  It was later shown that the essence of this
effect could be described by the two lowest energy eigenstates
corresponding to the symmetric $|S\rangle$, and antisymmetric
$|A\rangle$, states of the ground state doublet.  This approximation is
valid if the drive strength, drive frequency and energy splitting are
small compared to the energy spacing between the average energy of the
ground doublet and the higher energy levels \cite{Llo92}.  The two level
approximate Hamiltonian for this system written in a basis of left and
right states, $|L,R\rangle = (|S\rangle \mp |A\rangle)/\sqrt{2}$, 
is described by a bare tunneling system with an energy splitting $A_{x}$
and a cosine driving force with amplitude $A_{z}$ and frequency
$\omega_{d}$,
\begin{equation}
\hat{H}(t) = \frac{A_x}{2} \hat{\sigma}_{x} + \frac{A_z}{2}
\cos(\omega_{d} t)\hat{\sigma}_{z}.
\end{equation} 
A complete suppression of tunneling occurs when the drive amplitude is
much larger than the energy splitting ($A_z \gg A_x$) and the ratio of the
drive amplitude to the drive frequency is equal to a root of the zero
order Bessel function \cite{Llo92}.  This absence of coherent oscillation
is due to the two Floquet states of the driven system becoming exactly
degenerate \cite{Gro93}. 

\begin{figure}
\scalebox{1}{\includegraphics{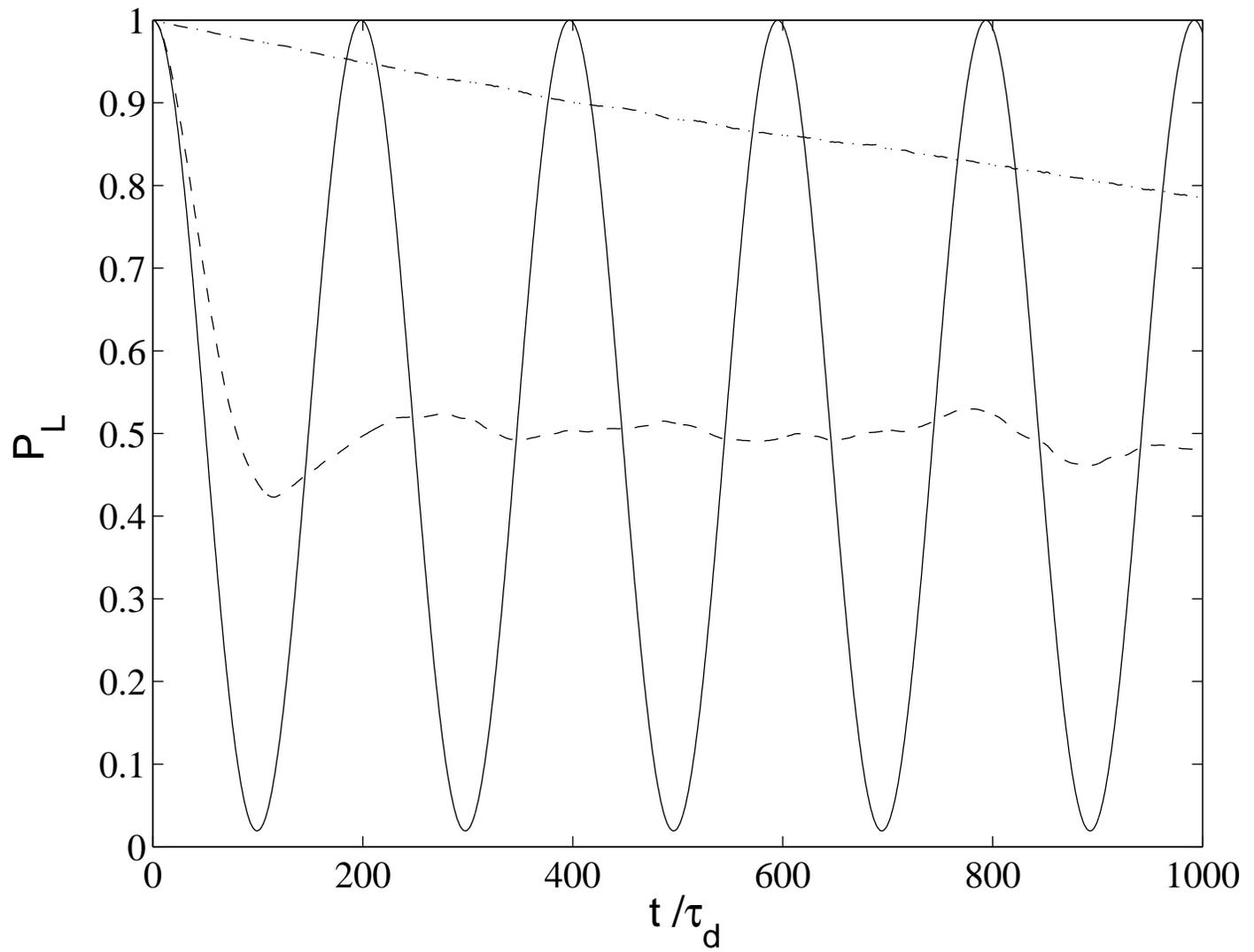}}
\caption{Fig.\ 1, J. Grondalski, Phys. Rev. E}
\label{fig1}
\end{figure}

A counterintuitive effect \cite{Gro93,Dit93,Koh98} occurs when dissipation
is added to this system.  This can be modeled by the addition of a
stochastic $\delta$-function kick, which is applied periodically
\cite{Gro93}.  When the ratio of the amplitude to the frequency of the
periodic driving field is near but not equal to the Bessel root condition
in the absence of noise, coherent oscillations between the left and right
wells proceed at a modified tunneling rate corresponding to the
quasienergy splitting (See~Fig.\ \ref{fig1}).  At first, the effect of
noise is to destroy coherent oscillations causing the system to evolve
into a 50/50 mixture of left and right wells.  However, increasing the
noise strength appears to have a stabilizing effect on coherent
suppression of tunneling \cite{Gro93,Dit93} causing the system to remain
in the well corresponding to its initial state.  This seems surprising
because the condition for coherent suppression of tunneling depends on the
state of the system being in a stationary superposition of the two
degenerate Floquet states for all times whereas dissipation generally
tends to destroy quantum coherence.  However, we will demostrate that this
localization effect can be understood intuitively as arising from strong
damping of the phase of oscillation.  This has been previously pointed out
by Makarov \cite{Mak93} using a semiclassical quantization scheme to
establish a correspondence between a classically driven two-level system
and the spin-boson problem.  Standard treatments of dissipation based on
the master equation of a reduced density operator after tracing over a
large reservoir \cite{Dit93,Mak93,Mak95} can obscure the physical content
of this localization phenomena.  Our treatment uses a simplified model
based on a damping due to phase diffusion from random classical kicks. 
This model corresponds to the regime of weak coupling to a classical noise
reservoir \cite{Leg87,Mak93,Mak95}.  From this we derive a detailed and
exact expression for the tunneling oscillations that gives a clear
interpretation of this phenomena that excludes any cooperative effects
between the noise and the driving field. 

Dissipation in driven two level systems has been extensively studied
\cite{Gro93,Dit93,Koh98,Mak93,Mak95,Wei99,Dak94,Gri95,Gri96a,Mak97,Neu97,Gri97,For99,Sto99}
because this model provides a realistic paradigm for understanding a
variety of physical systems.  For example this model can be used to
describe charge oscillations in semiconductor double wells
\cite{Bav92,Dak93a,Dak93b}, magnetic flux dynamics in superconducting
quantum interference devices (SQUIDs) \cite{Han91}, condensed phase
electron and proton reactions \cite{Mor93,Fle90,Ben94} and strong field
spectroscopy \cite{Cuk98,Mor99}.  Recently, quantum tunneling over
mesoscopic distances and the generation of mesoscopic superposition states
has been realized in an optical lattice of double wells \cite{Hay00}. 
This system is ideally suited for the study of the driven double well in
the presence of dissipation given the large degree of experimental control
one has over the system \cite{Deu98}.  For example the energy barrier and
energy assymmetry can be dynamically controlled through laser beam
configuration and externally applied magnetic fields.  Furthermore, this
system can operate in an essentially dissipation free environment when
the lattice lasers are sufficiently far detuned from the atomic resonance.
Dissipation can then be reintroduced into the system in the form of
well-controlled fluctuations in the potential \cite{Deu98,Deu00}. 

In this paper we replace the cosine drive field with a periodic square
wave drive field.  Our Hamiltonian reads,
\begin{eqnarray}
\hat{H}(t) & = & \frac{A_x}{2} \hat{\sigma}_{x} + \frac{A_z}{2}
\Lambda(\omega_{d} t)\hat{\sigma}_{z}, \nonumber \\*
&  & \\*
\Lambda(\omega_{d} t) & = & \left\{ \begin{array}{ll}
+1 & \mbox{  for   } 
0 \leq \mbox{mod}_{\tau_d}(t) < \frac{\tau_d}{2}\\
-1 & \mbox{  for   } 
\frac{\tau_d}{2} \leq \mbox{mod}_{\tau_d}(t) < \tau_d,
\end{array} \right. \nonumber
\end{eqnarray}
where $\tau_{d}=\frac{2 \pi}{\omega_{d}}$ is the drive period.  This drive
has similar qualitative features to the sinusoidal force.  For instance,
the condition for coherent suppression of tunneling occurs when
the ratio of the amplitude to the frequency of the drive is equal to an
integer rather than a root of the zero order Bessel function.  However,
this Hamiltonian has the advantage that one can derive analytic results
without making approximations that are common when analyzing sinusoidal
drives.  The paper is organized as follows.  In Sec.\ II we present an
effective magnetic field formalism in the Floquet basis showing how one
can describe the tunneling system with a set of Bloch equations associated
with spin precession about a fictitous static magnetic field.  In Sec.\
III we show how one can use a phase diffusion model to describe the
dissipation in terms of dephasing terms added to the Bloch equations and
we present results that compare numerical simulations to analytic
expressions.  In Sec.\ IV we summarize our results.

\section{The effective magnetic field in the Floquet basis}

\begin{figure}
\scalebox{1}{\includegraphics{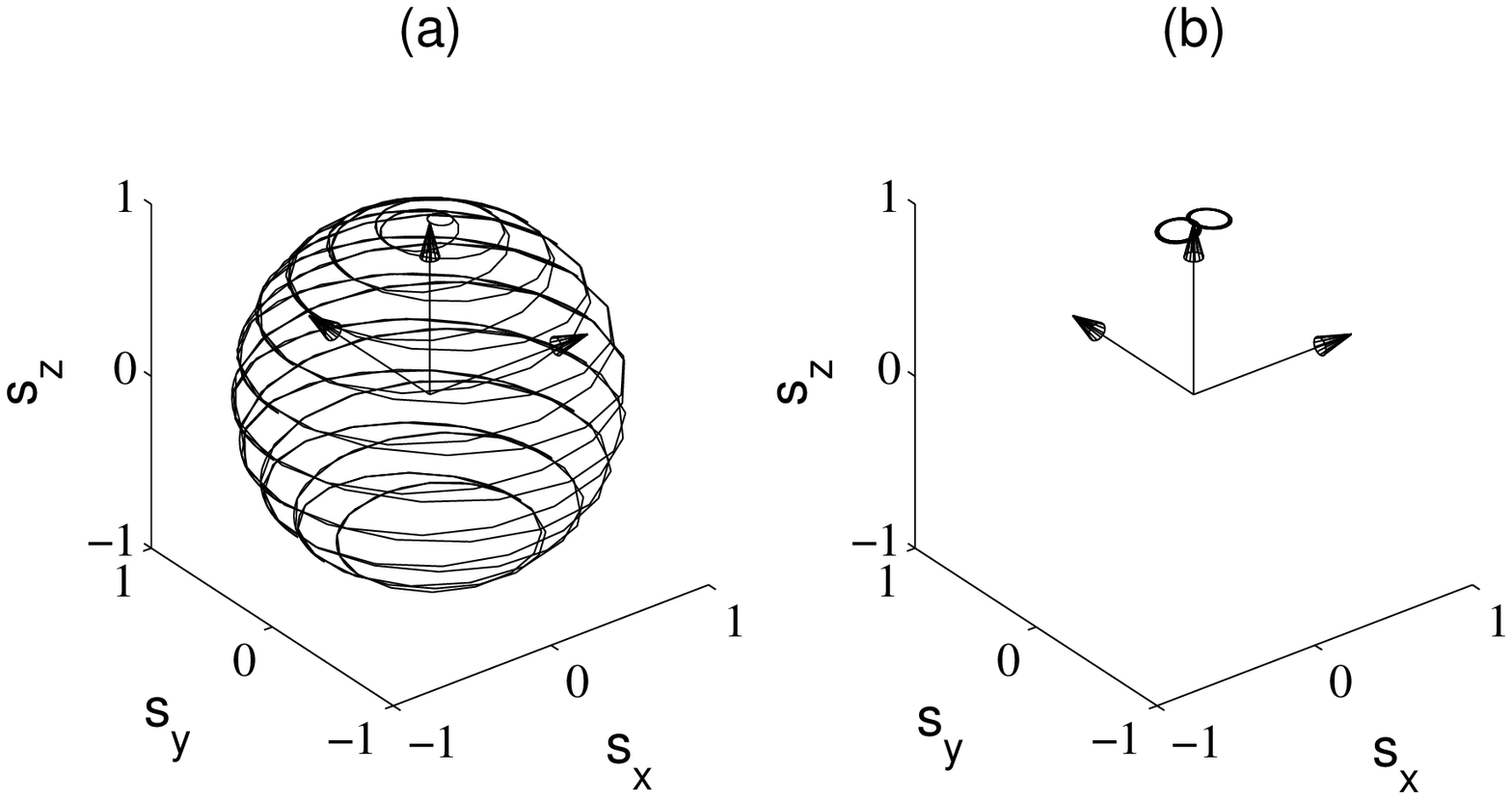}}
\caption{Fig.\ 2, J. Grondalski, Phys. Rev. E}
\label{fig2}
\end{figure}

As discussed above, we simplify the problem of a tunneling wave packet in
a double well by considering the problem restricted to the two lowest
energy eigenstates.  Because this is a two-level system one can visualize
the dynamics geometrically on the Bloch sphere with spin up and spin down
along the quantization axis $\mathbf{\hat{z}}$ corresponding
to the localized $|L\rangle$ and $|R\rangle$ states respectively.
According to the bare Hamiltonian, the first term in
Eq.\ (2), tunneling on the Bloch sphere is pictured as the Larmour
precession of a Bloch vector, $\mathbf{s}$, about a static magnetic
field in the $\mathbf{\hat{x}}$ direction with a frequency 
$\omega_{L}=A_{x}$ ($\hbar=1$).  When a drive is added to the tunneling
system, such as the square wave drive given in Eq.\ (2), the
Hamiltonian becomes time dependent.  In general, a drive term causes 
trajectories to explore the entire Bloch sphere in a complicated fashion
(See Fig.\ \ref{fig2}a).  However, when the ratio of the amplitude
to the frequency of the drive is an integer and $A_z \gg A_x$, one finds
that the trajectories on the Bloch sphere form closed ``loops'' near the
top (or bottom) of the Bloch sphere which correspond to the quantum system
remaining localized in the $|L\rangle$ (or $|R\rangle$) state (See Fig.\
\ref{fig2}b).  This is the coherent suppression of tunneling.  Note that
if the condition $A_z \gg A_x$ 
(typically $A_z/A_x \gtrsim 5$) is not satisfied then
the system can still form closed trajectories but they do not stay
``near'' the top or bottom of the Bloch sphere and hence are not 
localizing.

According to the Floquet theorem \cite{Shi65,Gri98}, a system with a 
Hamiltonian that is periodic in time, $\hat{H}(t + \tau_d) = \hat{H}(t)$, 
has solutions to the Schr\"{o}dinger equation, $|\varepsilon_{j}\rangle$,
which are eigenfunctions of the single period time propagator or Floquet
operator, ($\hbar=1$)
\begin{equation}
\hat{U}(\tau_{d},0) = \hat{\mathcal{T}}
\exp\Bigl(-i \int_{0}^{\tau_{d}} dt \hat{H}(t)\Bigr)
\end{equation}
where $\hat{\mathcal{T}}$ denotes the time-ordering operator.  The
eigenvalues of this propagator are given by
\begin{equation}
\hat{U}(\tau_{d},0) |\varepsilon_{j}\rangle =
\exp(-i \varepsilon_{j} \tau_{d}) |\varepsilon_{j}\rangle,
\end{equation}
where $\varepsilon_{j}$ is a quasienergy which belongs to a family of
quasienergies such that $\varepsilon_{j} + k \omega_{d}$ (where $k$ is an
integer) belongs to the same physical state.  Quasienergies can be   
uniquely defined by requiring them to continuously approach the energies
of the time independent Hamiltonian as the periodic part vanishes.
Because the Floquet states are stationary states at integer multiples of
$\tau_{d}$, we restrict time to these discrete values.   In this
stroboscopic picture of the dynamics, the Floquet states evolve like 
energy eigenstates for a time-independent Hamiltonian with the
quasienergies playing the role of energies.

\begin{figure}
\scalebox{1}{\includegraphics{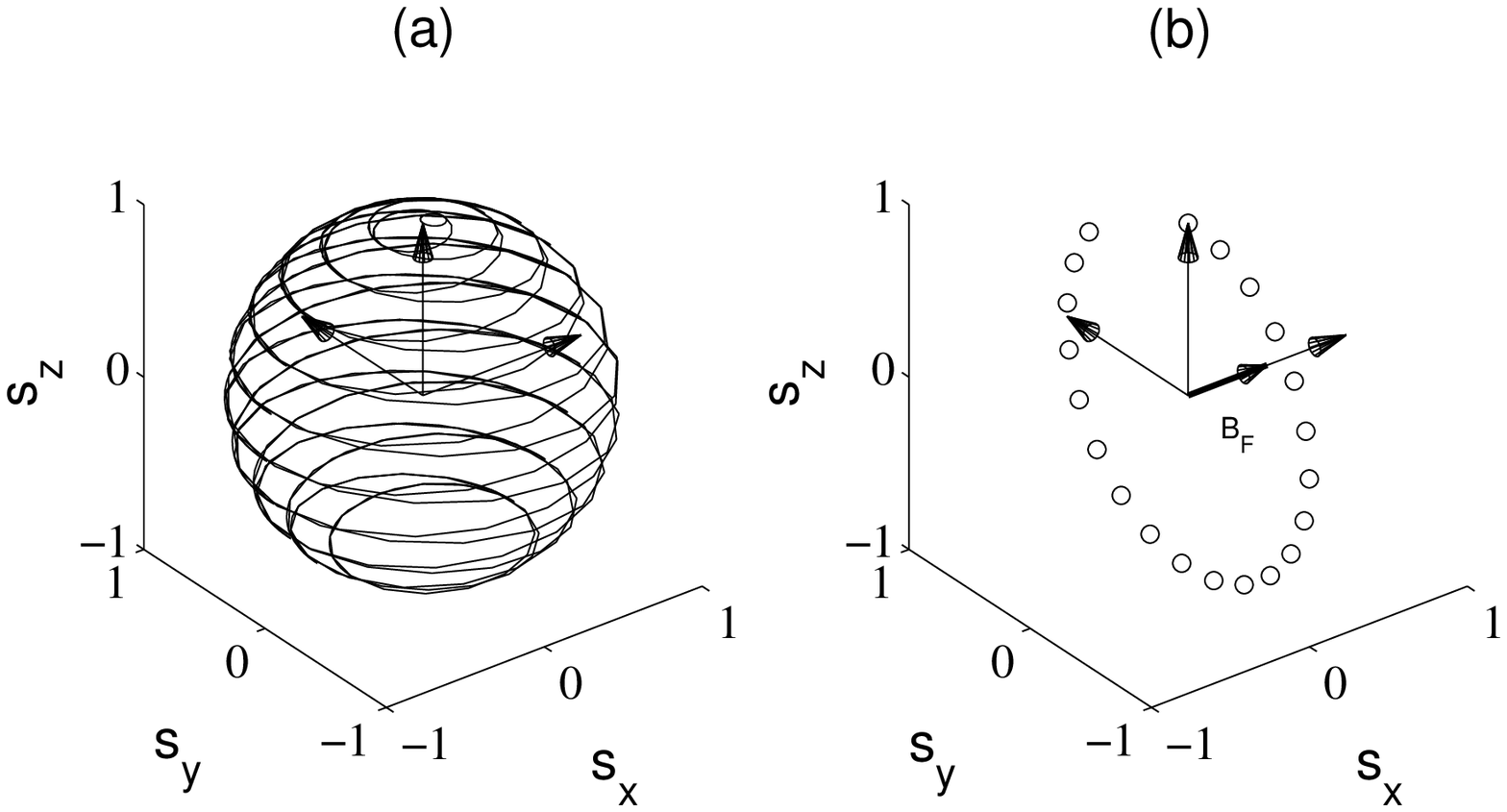}}
\caption{Fig.\ 3, J. Grondalski, Phys. Rev. E}
\label{fig3}
\end{figure}

For a two-level system we denote the Floquet states
$\{|\varepsilon_{-}\rangle,|\varepsilon_{+}\rangle\}$.  
Writing the single period time propagator in the Floquet basis and
neglecting global phase factors gives,
\begin{equation}
\hat{U}(\tau_{d},0) = 
\hat{1}\cos\Bigl(\frac{\Delta \varepsilon \tau_d}{2}\Bigr)
- i( \mathbf{\hat{\vec{\sigma}}} \cdot \mathbf{\hat{e}}_{F})
\sin\Bigl(\frac{\Delta \varepsilon \tau_d}{2}\Bigr),
\end{equation}
where $\Delta \varepsilon = \varepsilon_{+}-\varepsilon_{-}$ is the
quasienergy splitting and the quantization direction,
$\mathbf{\hat{e}}_F$, is chosen along $|\varepsilon_{+}\rangle$.  We see
that the application of the single period time propagator can be thought
of as a rotation about an effective magnetic field, 
$\mathbf{B}_{F}=\Delta\varepsilon \; \mathbf{\hat{e}}_F$.
Because the single period time propagator commutes with itself at integer
multiples of the drive period, the unitary time propagator after $\ell$
drive periods is found from $\ell$ applications of the single period time
propagator, $\hat{U}(\ell \tau_d,0) = \hat{U}(\tau_{d},0)^{\ell}$ 
($\ell=0,1,2,\cdots$).  The end result of utilizing this representation is
that one can replace the continuous dynamics of a time dependent
Hamiltonian with a discrete dynamics associated with a time independent
Hamiltonian described by spin precession about a fictitous static magnetic
field (See Fig\ \ref{fig3}). This transformation is not an approximation,
like the well known rotating wave approximation; the two pictures are in
exact agreement at these discrete times.  However, the time resolution of
the Bloch trajectories is fundamentally limited by the drive period
because this quasistatic picture does not contain information about time
scales shorter than the drive period.

For the specific case of the square wave drive the time dependent
Hamiltonian can be decomposed into two time independent Hamiltonians,
$\hat{H}_{\pm}=\frac{A_x}{2}\hat{\sigma}_{x} \pm
\frac{A_z}{2}\hat{\sigma}_{z}$.  According to Eq.\ (2), the single
period unitary operator can be easily contructed by multiplying the two
half period unitaries,
\begin{subequations}
\label{allequations}
\begin{eqnarray}
\hat{U}(\tau_d,0) & = & \exp\Bigl(-i \hat{H}_{-} \frac{\tau_d}{2}\Bigr)
\exp\Bigl(-i \hat{H}_{+} \frac{\tau_d}{2} \Bigr) \label{equationa} \\
& = &  \Bigl( 1-2\sin^{2}(\frac{\phi}{2})\sin^{2}(\theta) \Bigr)
\hat{1} \nonumber \\
&   &
-\Bigl( 2 i \sin(\frac{\phi}{2})\cos(\frac{\phi}{2})\sin(\theta) \Bigr)
\hat{\sigma}_x \nonumber \\
&   &
+\Bigl( 2 i \sin^{2}(\frac{\phi}{2})\sin(\theta)\cos(\theta) \Bigr)
\hat{\sigma}_y, \label{equationb}
\end{eqnarray}
\end{subequations}
where $\phi = \sqrt{A_x^2 + A_z^2}\; \tau_d$ and 
$\tan(\theta) = A_z/A_x$.  The effective magnetic field for the square
wave drive can be found analytically by comparing Eq.\ (6b) to
\begin{equation}
\exp\Bigl(-i(\mathbf{\hat{\vec{\sigma}}\cdot\hat{n}})
\frac{\Delta \varepsilon \tau_d}{2}\Bigr) =
\hat{1}\cos\Bigl(\frac{\Delta \varepsilon \tau_d}{2}\Bigr) -
i(\mathbf{\hat{\vec{\sigma}}\cdot\hat{n}})
\sin\Bigl(\frac{\Delta \varepsilon \tau_d}{2}\Bigr),
\end{equation}
and solving for the four quantities $\{\Delta \varepsilon,n_x,n_y,n_z\}$.
With a proper coordinate transformation, we find that the effective
magnetic field for a square wave drive can always be viewed to be in the
$\mathbf{\hat{x}}$ direction,  $\{n_{x}=1,n_{y}=0,n_{z}=0\}$, and the
quasienergy
splitting is given by
\begin{equation}
\Delta \varepsilon = \frac{2}{\tau_{d}} 
\cos^{-1}\Bigl( 1 - 2\sin^{2}(\phi/2) \sin^{2}(\theta) \Bigr).
\end{equation}
One can verify that the effective magnetic field vanishes 
(Bloch vector trajectories form closed ``loops'' on the Bloch sphere)
when the second term in Eq.\ (8) vanishes $\bigl( \sin(\phi/2)=0 \bigr)$
giving the condition $\sqrt{A_x^2 + A_z^2}/\omega_d = j$
($j$~is~an~integer).  When $A_z \gg A_x$ this reduces to the condition,
$A_z/\omega_d=j$, the same condition stated in Sec.\ I for coherent
suppression of tunneling by a square wave drive.  

\section{Dissipation}  

We next add dissipation to the system with stochastic periodic
$\delta$-function kicks \cite{Gro93}.  Equation~(2) is modified to read,
\begin{equation}
\hat{H}(t) = \frac{A_x}{2} \hat{\sigma}_{x} +
\biggl(\frac{A_z}{2} \Lambda(\omega_d t) +
\sum_{\ell}\xi_{\ell} \; \delta(t - \ell \tau_d) \biggr)
\hat{\sigma}_{z},
\end{equation}
where $\xi_{\ell}$ is a discrete random variable governed by a
distribution, $P(\xi)$, defined by a mean, $\langle\xi\rangle$, and
variance, $\langle \xi^2 \rangle=\sigma^2$.  This model can be related to
the widely studied spin-boson model in the weak coupling limit where
dissipation is accomplished through coupling to a reservoir of harmonic
oscillators at temperature $T$.  We identify the variance with 
$\sigma^2~=~4\hbar^{-1}J(\Delta \varepsilon) 
\coth(\hbar \Delta \varepsilon /2 k_{b} T)$, where $J(\omega)$ is the
``spectral function'' of the reservoir \cite{Leg87,Mak93}.

Without loss of generality we take $\langle\xi\rangle=0$ since a non-zero
mean can be removed by an appropriate redefinition of the single period
time propagator
\begin{equation}
\hat{U}(\tau_{d},0) \rightarrow 
\exp\Bigl(-i (\mathbf{\hat{\vec{\sigma}} \cdot \hat{z}})
\frac{\langle\xi\rangle}{2} \Bigr)
\; \hat{U}(\tau_{d},0).
\end{equation}
The average dynamics are essentially independent of the specific shape of
the distribution, $P(\xi)$, because of the central limit theorem.  On
the Bloch sphere, the dynamics can be viewed as single period time
rotation about the effective magnetic field followed by a stochastic
rotation about the $\mathbf{\hat{z}}$ axis by an angle $\xi_{\ell}$,
\begin{equation}
\mathbf{\vec{s}}(t) = \Bigl( \mathcal{R}_{\mathbf{\hat{z}}}(\xi_{\ell})
\mathcal{R}_{\mathbf{\hat{F}}}(\Delta\varepsilon \tau_d)\Bigr)^{\ell} \:
\mathbf{\vec{s}}(0).
\end{equation}
Taking the rotations about $\mathbf{\hat{z}}$ is completely general;
kicks about other directions can be implemented with an approprate unitary 
transformation.  The mean trajectory is computed by averaging over many
realizations of the stochastic evolution.  To see how the averaging
effects the dynamics, consider the average $x$ and $y$ components of the
Bloch vector, $\langle s_{j}(t) \rangle$ ($j={x,y}$). After $\ell$ periods
of the drive, the random rotations about $\mathbf{\hat{z}}$ cause the
components to execute a random walk in the
$\mathbf{\hat{x}}-\mathbf{\hat{y}}$ plane.  Accordingly, the angle in the
$\mathbf{\hat{x}}-\mathbf{\hat{y}}$ plane obeys a diffusion equation and
the mean $x$ and $y$ components decay as $\langle s_{j}(\ell
\tau_d)\rangle=\exp(-\frac{\ell \tau_d}{T_{2}})$, where the characteristic
dephasing time is $T_{2} = 2/\sigma^2$ \cite{Ste84}. These decay terms due
to dephasing can be included in the Bloch equations,
\begin{equation}
\begin{array}{lll}
\dot{s_x} &=& \Bigl( B_y\;s_z - B_z\;s_y \Bigr) - \frac{1}{T_2} \; s_x\\
\dot{s_y} &=& \Bigl( -B_x\;s_z + B_z\;s_x \Bigr) - \frac{1}{T_2} \; s_y \\
\dot{s_z} &=& \Bigl( B_x\;s_y - B_y\;s_x \Bigr).
\end{array}
\end{equation}
Even though these are differential equations associated with a continuous
time variable, there is an implicit understanding that the solution to
these equations agree with the real system only at multiple integers of
the drive period, $t = \ell \tau_d$.

As an example, we show how these Bloch equations can be used to
derive the localization effect of dissipation caused by stochastic
rotations about $\mathbf{\hat{z}}$ that was discussed in Sec.\ I.  We take
the initial state to be localized in the left well.  Because the
effective magnetic field for the periodic square drive is only in the
$\mathbf{\hat{x}}$ direction, the dynamics take place completely in the
$\mathbf{\hat{y}}-\mathbf{\hat{z}}$ plane.  This reduces the number of
differential equations that need to be solved to two which can be written
as a single second order differntial equation,
\begin{equation}
\ddot{s_z} + \frac{1}{T_2} \; \dot{s_z} + 
(\Delta \varepsilon)^2 \; s_z = 0
\end{equation}
with the initial conditions, $\{s_{z}(0)=1,\dot{s_z}(0)=0\}$.  The
solution to this equation is the familiar damped harmonic oscillator which
can be solved in three regions: underdamped
($\Delta\varepsilon^2 > \beta^2$), critically damped 
($\Delta\varepsilon^2 = \beta^2$), and overdamped
($\Delta\varepsilon^2 < \beta^2$), where $\beta = 1/2T_2$,
\begin{equation}
s_{z}(t) = \left\{
\begin{array}{lr}
\frac{1}{\cos(\delta)} e^{-\beta t}
\cos\Bigl(\sqrt{\Delta \varepsilon^2 - \beta^2} \; t - \delta\Bigr), &
\Delta\varepsilon^2 > \beta^2\\
e^{-\beta t}(1 + \beta t), &
\Delta\varepsilon^2 = \beta^2\\
e^{-\beta t}\biggl(\frac{\sqrt{\beta^2 - \Delta\varepsilon^2}+
\beta}{2\sqrt{\beta^2 - \Delta \varepsilon^2}}
\exp\Bigl(\sqrt{\beta^2 - \Delta \varepsilon^2}\: t \Bigr) +
\frac{\sqrt{\beta^2 - \Delta\varepsilon^2}-
\beta}{2\sqrt{\beta^2 - \Delta \varepsilon^2}}
\exp\Bigl(-\sqrt{\beta^2 - \Delta \varepsilon^2}\: t \Bigr) \biggr), &
\Delta\varepsilon^2 < \beta^2,
\end{array} \right.
\end{equation}
where $\tan(\delta) = \beta/\sqrt{\Delta\varepsilon^2 - \beta^2}$.

For a particle intially in the $|L\rangle$ state and for parameters of the
Hamiltonian, $A_x=0.1$, $A_z=1.1$, and $\omega_d = 1$, we perform
numerical simulations according to the procedure described by Eq.\ (11)
and average over 500 Bloch trajectories.  We compare plots of the
probability $P_{L}$ to be in the $|L\rangle$ state vs. time for
these simulations with results obtained from Eq.\ (14) and find very good
agreement (Fig.\ 4~a--c).  However, the model eventually breaks down when
$T_{2}<\tau_d$ (Fig.\ 4~d).  In this regime the system is decaying faster
than the oscillation frequency of the drive, the fundamental time
resolution for working in the Floquet basis as discussed in Sec.\ II.

We are now in a position to explain the localization effect of dissipation
that was presented in Sec.\ I.  For small $\sigma$, dissipation causes the
quantum oscillations to decay on a time scale $T_{2}$.  Increasing
$\sigma$ (decreasing $T_{2}$) causes the oscillations to decay on a
shorter time scale until they reach a critical damping where the coherent
oscillations are completely destroyed.  Increasing $\sigma$ further causes
the system to be overdamped with a dephasing time given approximately by
$1/(T_{2} \Delta\varepsilon^2)$.  In this regime increasing
$\sigma$ (decreasing $T_{2}$) causes the system to decay on a {\em longer}
time scale.  The classical noise localizes the state by destroying the
phase coherence necessary for transitions between the left and right
wells and not by any cooperative effect between the noise and the driving
field.  This can be seen from the fact that this localization effect
would still occur, for a small enough bare tunneling splitting, even if
the drive were completely eliminated!  The reason this localization effect
is stronger ``near'' the condition for coherent suppression of tunneling
is because the dephasing time for this overdamping is longer at smaller
quasienergy splittings.  This explains why dissipation appears to 
``stabilize'' coherent suppression of tunneling.  We have performed
simulations for the two-level system driven by a cosine drive and have
found similar results, the only difference being that one must compute the
effective magnetic field numerically. 

\section{Conclusion}

In this article we studied the localization effect of dissipation caused
by random periodic $\delta$-function ``kicks'' on a driven two-level
system.  Utilizing a Floquet formalism we found that this type of
dissipation could be modeled with a set of Bloch equations including phase
damping terms.  For dephasing times longer than the drive period, these
equations yield dynamics that agree with the periodically kicked system at
discrete times.  We considered a square wave drive in this article because
it gives the same qualtative results as the widely studied cosine drive,
but it is simpler to study.  We found that the tunneling oscillations obey
the equations of a damped harmonic oscillator and that this localization
effect corresponds to overdamped tunneling oscillations.  Thus, this
stabilization of coherent suppression of tunneling is due simply to strong
phase damping of tunneling oscillations, which effectively projects the
system into the pointer-basis set by the interaction of the system with
the noisy environment \cite{Zur82}, rather than a cooperative effect
between the noise and the driven system.  The simplified model studied in
this article allowed one to gain physical intuition about the
stabilization of coherent suppression of tunneling in the presense of
noise and may prove useful in the study of other driven dissipative
quantum systems in the weak coupling limit such as dynamical localization
in a lattice \cite{Dun86,Hol92,Bha99} or quantum stochastic resonance
\cite{Gri96b,Lof94}.  The utility of the square wave drive over the
commonly studied sinusoidal drives may find wider application in more
general treatments of driven quantum dissipative systems.

\begin{acknowledgments}

This research was supported by NSF Grant No. PHY-9732456.

\end{acknowledgments}

\bibliography{coherent}

\pagebreak

Fig.\ 1:  The probability to be in the left well, $P_{L}$, vs. time,
$t/\tau_{d}$.  For the central cosinusoidal curve (solid line) $\sigma=0$,
where $\sigma$ is the root mean variance of the probability distribution
which governs the discrete random variable, we have coherent oscillations
between the $|L\rangle$ and $|R\rangle$ states at a modified tunneling
rate corresponding to the quasienergy splitting.  In the central rapidly
decaying curve (dashed), $\sigma=0.25$, stochastic $\delta$-function kicks
rapidly destroy coherent oscillations.  In the upper, slowly decaying
curve (dash-dotted), $\sigma=2.5$, the coherent suppression of tunneling
has been partially restored by the increased noise strength.  Compare this
to Fig.\ 7 b on pg.~243 in Ref.\ \cite{Gro93} and Fig.\ 2 on pg.~9 in
Ref.\ \cite{Dit93}. 

Fig.\ 2:  The representation of coherent suppression of tunneling on the  
Bloch sphere for a two-level system initally in the $|L\rangle$ state:
a) Trajectories off the condition of coherent suppression of tunneling 
wind up and down the Bloch sphere.  b) When the condition for coherent
suppression of tunneling is met trajectories form closed ``loops'' near
the top ($|L\rangle$) of the Bloch sphere.

Fig.\ 3:  Analyzing the system in a Floquet basis allows one to replace
the (a) complicated continuous dynamics associated with a time dependent
Hamiltonian with a (b) simple discrete dynamics associated with
Larmour precession about a fictitous static magnetic field,
$\mathbf{B}_F$.  The two pictures agree exactly at these discrete times
corresponding to integer multiples of the drive period.

Fig.\ 4:  Comparison of numerical simulations (dotted line) to analytic
expression (solid line).  Each graph plots the probability $P_{L}$ to be
in the left well vs.\ time, $t/\tau_{d}$, for the parameters $A_x=0.1,
A_z=1.1$, and $\omega_d = 1$. There is excellent agreement for cases:
a) underdamped, ($\sigma = 0.20, T_2 = 50 \tau_d$), b) critically damped
($\sigma = 0.68, T_2 = 4.28 \tau_d$), and c) overdamped
($\sigma = 1.41, T_2 = 1 \tau_d$).  The model breaks down:
d) ($\sigma = 2.5, T_2 = .32 \tau_d$) when $T_2 < \tau_d$.

\pagebreak

\clearpage

\clearpage

\clearpage

\begin{figure}
\scalebox{1}{\includegraphics{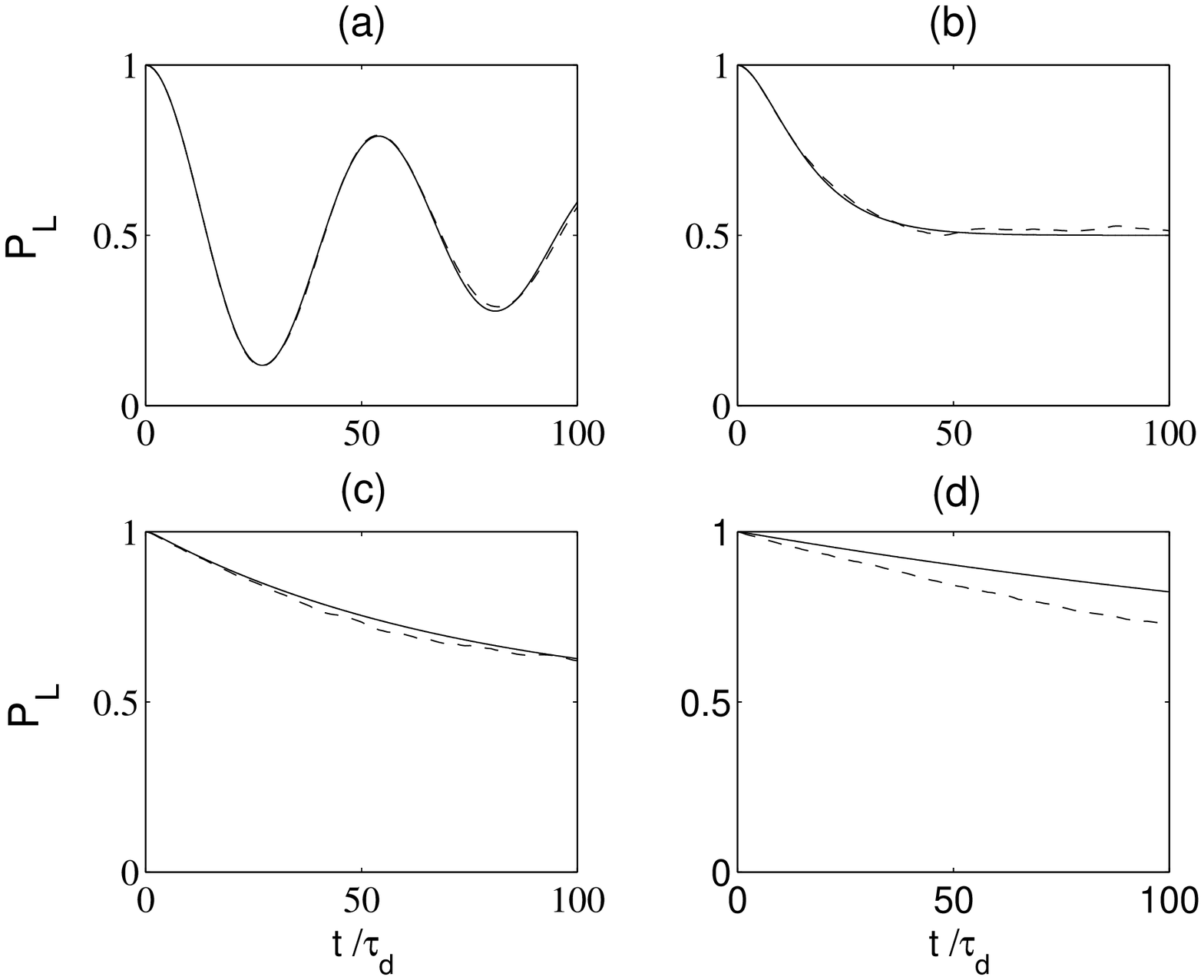}}
\caption{Fig.\ 4, J. Grondalski, Phys. Rev. E}
\label{fig4}
\end{figure}

\end{document}